\documentclass[twocolumn]{aa}
\usepackage{graphicx,color}
\usepackage{txfonts}
\begin{document}
\def\grs1915{GRS~1915$+$105}
\newcommand{\nh}{N_{\rm H}}
\newcommand{\wfe}{W_{{\rm K}\alpha}}
\newcommand{\efe}{E_{{\rm K}\alpha}}
\newcommand{\sfe}{\sigma_{{\rm K}\alpha}}
\newcommand{\ee}{$e^\pm$}
\newcommand{\g}{$\gamma$}
\newcommand{\lh}{\ell_{\rm h}}
\newcommand{\ls}{\ell_{\rm s}}
\newcommand{\lth}{\ell_{\rm th}}
\newcommand{\lnth}{\ell_{\rm nth}}
\newcommand{\sax}{{\it Beppo\-SAX}}
\newcommand{\source}{GRS 1915+105}
\newcommand{\msun}{{\rm M}_\sun}
\newcommand{\xte}{{\it RXTE}}
\newcommand{\pca}{{\it RXTE}/PCA}
\newcommand{\asca}{{\it ASCA}}
\newcommand{\gro}{{\it CGRO}}
\newcommand{\asm}{{\it RXTE}/ASM}
\newcommand{\integral}{{\it INTEGRAL}}
   \title{Characterizing a new class of variability in GRS~1915$+$105
   with simultaneous {\it INTEGRAL/RXTE} observations}


   \author{D.C. Hannikainen
          \inst{1} 
        \and
	  J. Rodriguez\inst{2,3} \and
	  O. Vilhu\inst{1} \and
	  L. Hjalmarsdotter\inst{1} \and
	  A.A. Zdziarski\inst{4} \and
	  T. Belloni\inst{5} \and 
          J. Poutanen\inst{6} \and
	  K. Wu\inst{7} \and
          S.E. Shaw\inst{8,3} \and 
          V. Beckmann\inst{9,10} \and 
	  R.W. Hunstead\inst{11} \and
	  G.G. Pooley\inst{12} \and
	  N.J. Westergaard\inst{13} \and
	  I.F. Mirabel\inst{14} \and
	  P. Hakala\inst{1} \and
	  A. Castro-Tirado\inst{15} \and
	  Ph. Durouchoux\inst{2}
          }

   \offprints{D.C. Hannikainen: diana@astro.helsinki.fi}

   \institute{Observatory, PO Box 14, FIN-00014 University of
              Helsinki, Finland
        \and
           Centre d'Etudes de Saclay, DAPNIA/Service
           d'Astrophysique (CNRS FRE 2591), Bat. 709, Orme des
           Merisiers, Gif-sur-Yvette Cedex 91191, France
        \and
	   {\it INTEGRAL} Science Data Center, Chemin d'\'Ecogia 16,
           CH-1290 Versoix, Switzerland
	 \and
	      Nicolaus Copernicus Astronomical Center, Bartycka 18,
              00-716 Warszawa, Poland
	  \and
	   INAF -- Osservatorio Astronomico di Brera, via E. Bianchi
              46, 23807 Merate (LC), Italy
	 \and     
	     Astronomy Division, University of Oulu, PO Box 3000,
              90014 Finland
	 \and
	      MSSL, University College London, Holmbury St. Mary,
              Surrey, RH5 6NT, UK
	 \and 
	    Dept. of Physics and Astronomy, University of Southampton, 
              Southampton SO17 1BJ, UK
	 \and 
              NASA Goddard Space Flight Center, 
              Code 661, Greenbelt, MD 20771, USA
         \and
              Joint Center for Astrophysics, Department of Physics, 
              University of Maryland, Baltimore County, Baltimore, MD
              21250, USA
	  \and
	     School of Physics, University of Sydney, NSW 2006,
             Australia
	  \and 
	     Astrophysics Group, Cavendish Laboratory, University of 
             Cambridge, Madingley Road, Cambridge CB3 0HE, UK
         \and
	   Danish Space Research Institute, Juliane Maries Vej 30,
             2100 Copenhagen {\O}, Denmark
	  \and
	   European Southern Observatory, Santiago 10, Chile
	  \and
	    Instituto de Astrof\'{\i}sica de Andaluc\'{\i}a
              (IAA-CSIC), PO Box 03004, 18080 Granada, Spain
}

   \date{Received ; Accepted}

   \abstract{We report on the analysis of 100~ks {\it INTEGRAL}
   observations of the Galactic microquasar \grs1915. 
   We focus on {\it INTEGRAL} Revolution number 48 when the 
   source was found to exhibit a new type of 
   variability as preliminarily reported in Hannikainen 
   et al. (2003).
   The variability pattern, which we name $\xi$, 
   is characterized by a pulsing behaviour, consisting 
   of a main pulse and a shorter, softer, and smaller amplitude
   precursor pulse, on a timescale of 5 minutes
   in the JEM-X 3--35~keV lightcurve. 
   We also present simultaneous {\it RXTE} data.
   From a study of the individual {\it RXTE}/PCA 
   pulse profiles we find that the rising phase
   is shorter and harder than the declining phase, which is opposite to 
   what has been observed in other otherwise similar variability
   classes in this source.
   The position in the colour-colour diagram 
   throughout the revolution corresponds to State A (Belloni et
   al. 2000) but not to any previously known variability class.
   We separated the {\it INTEGRAL} data into two subsets covering the
   maxima and minima of the pulses and fitted the resulting two 
   broadband spectra
   with a hybrid thermal--non-thermal Comptonization model.
   The fits show the source to be in a soft state characterized by a
   strong disc component below $\sim$6~keV and Comptonization by both
   thermal and non-thermal electrons at higher energies. 
   
   }

\authorrunning{Hannikainen et al.}
\titlerunning{GRS~1915$+$105}

   \maketitle
%

\section{Introduction}

GRS~1915$+$105 has been extensively observed at all wavelengths ever
  since its discovery.
It was originally detected as a hard X-ray source with
  the WATCH all-sky monitor on the {\it GRANAT} satellite (Castro-Tirado,
  Brandt  \& Lund 1992). 
Apparent superluminal ejections have been observed from
  GRS~1915$+$105 several times with the Very Large Array 
  in 1994 and 1995 (Rodr\'\i guez \& Mirabel 1999) 
  and in 1997 with the Multi-Element Radio-Linked
  Interferometer Network (Fender et al. 1999).
Both times the true ejection velocity was calculated to be $>0.9c$.
In addition to these superluminal ejections at the arcsec scale,
  \grs1915 has also exhibited a compact radio jet resolved at
  milli-arcsec scales corresponding to a length of a few tens of AU
  (Dhawan, Mirabel \& Rodr\'\i guez 2000; Fuchs et al. 2003).
Using the Very Large Telescope, Greiner et al. (2001)
  identified the mass-donating star to be of spectral type K-M III.
The mass of the black hole was deduced by Harlaftis \& Greiner (2004)
  to be $14.0\pm4.4\,{\rm M}_{\odot}$ in a binary orbit with the giant star
  of 33.5 days.
The Rossi X-ray Timing Explorer ({\it RXTE}) has observed GRS~1915$+$105
  since its launch in late 1995 and has shown the source to be highly
  variable on all timescales from milliseconds to months
  (see e.g. Belloni et al. 2000; Morgan, Remillard \& Greiner 1997).
Belloni et al. (2000) categorized the variability into twelve distinct
  classes which they labelled with Greek letters (two of
  those represented by semi-regular pulsing behaviour), and
  identified three distinct X-ray states: two softer
  states, A and B, and a harder state, C. 
The source has been detected up to $\sim$~700~keV during OSSE observations
  (Zdziarski et al. 2001).
For a recent comprehensive review of \grs1915 see Fender \& Belloni
  (2004). \\
\indent The European Space Agency's INTErnational Gamma-Ray Astrophysical 
  Laboratory ({\it INTEGRAL}) is aimed at observing the sky between 
  $\sim$ 3~keV and 10~MeV (Winkler et al. 2003).
The {\it INTEGRAL} payload consists of two gamma-ray instruments
  -- the Imager on Board the {\it INTEGRAL} Spacecraft, IBIS
  (Ubertini et al. 2003), and 
   the Spectrometer on {\it INTEGRAL}, SPI (Vedrenne et al. 2003); two
   identical X-ray monitors -- the Joint European X-ray monitor, JEM-X
   (Lund et al. 2003); and an optical monitor (Mas-Hesse et al. 2003). \\
\indent GRS~1915$+$105 is being observed with {\it INTEGRAL}
  as part of the Core Program and also within the framework of an Open
  Time monitoring campaign.
Here we present the results of the first of our AO-1 Open
  Time observations which took place during {\it INTEGRAL} Revolution
  number 48.
The source was found to exhibit a new type of variability not
  documented in Belloni et al. (2000). 
Preliminary results were described elsewhere (Hannikainen et
  al. 2003).
We also have a simultaneous observing campaign with {\it RXTE} to
  concentrate on timing analysis (Rodriguez et al. 2004a).
The main goal of this {\it INTEGRAL/RXTE} program was to 
  study the disc-jet connection in \grs1915 and to study the
  variability classes in the hard X-ray regime.
As the source behaviour is very unpredictable, we were not able to 
  study the disc-jet connection in this particular case, nor any
  of the already known variability classes. 
Instead, (as we show below), the source exhibited a new type of 
  variability and we discovered a new source (IGR~J19140$+$0951)
  during the observation.
In this paper, we conduct a more detailed analysis of the high energy
  behaviour of \grs1915 during Revolution 48. 
In Section~\ref{datared} we discuss the observation and data reduction
  methods, while in Section~\ref{variability} we re-introduce the new
  type of variability. 
In Section~\ref{rxte} we present the {\it RXTE} results,
  while in Section~\ref{broadspec} we discuss the spectral analysis of the {\it
  INTEGRAL} data. 
Finally, in Section~\ref{concl} we present a summary of the paper.


\section{Observations and data reduction}\label{datared}

\begin{figure}
   \centering
   \includegraphics[width=9cm]{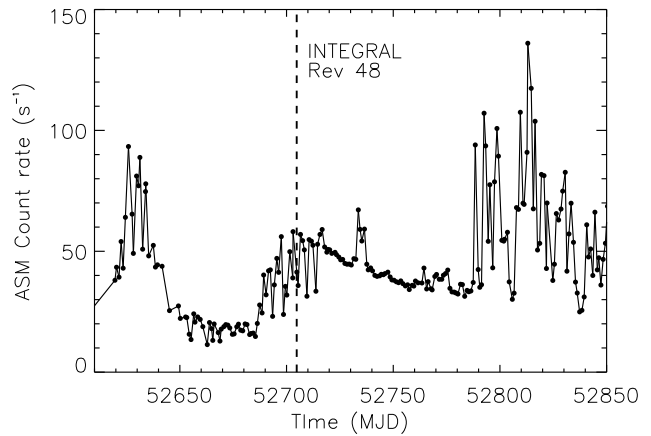}
      \caption{The {\it RXTE}/ASM one-day average 1.3--12~keV
              lightcurve showing the epoch of the {\it INTEGRAL}
              observation and the simultaneous pointed {\it RXTE}
              observation (dashed line).
              }
         \label{fig:asm}
   \end{figure}

As part of a monitoring programme which consists of six 100-ks
  observations, {\it INTEGRAL} observed \grs1915 for the first time for
  this campaign on 2003 Mar 6 beginning at UT 03:22:33 during {\it
  INTEGRAL's} Revolution 48 for 100~ks. 
  The {\it INTEGRAL} observations were undertaken using the hexagonal
  dither pattern (Courvoisier et al. 2003), which consists of 
  seven pointings around a nominal target location (1 source on-axis
  pointing, 6 off-source pointings, each $2^{\circ}$ apart, with each
  science window of 2200 s duration).
This means that \grs1915 was always in the field-of-view of all three
  X-ray and gamma-ray instruments (JEM-X, IBIS and SPI) throughout the
  whole observation. \\
\indent Figure~\ref{fig:asm} shows the {\it RXTE}/ASM one-day average
  1.3--12~keV lightcurve.
The vertical line indicates the date of the {\it INTEGRAL} and the simultaneous {\it
  RXTE} pointed observations.

\subsection{{\it INTEGRAL} data reduction}

\begin{figure*}[htbp]
   \centering
   \includegraphics[width=18cm]{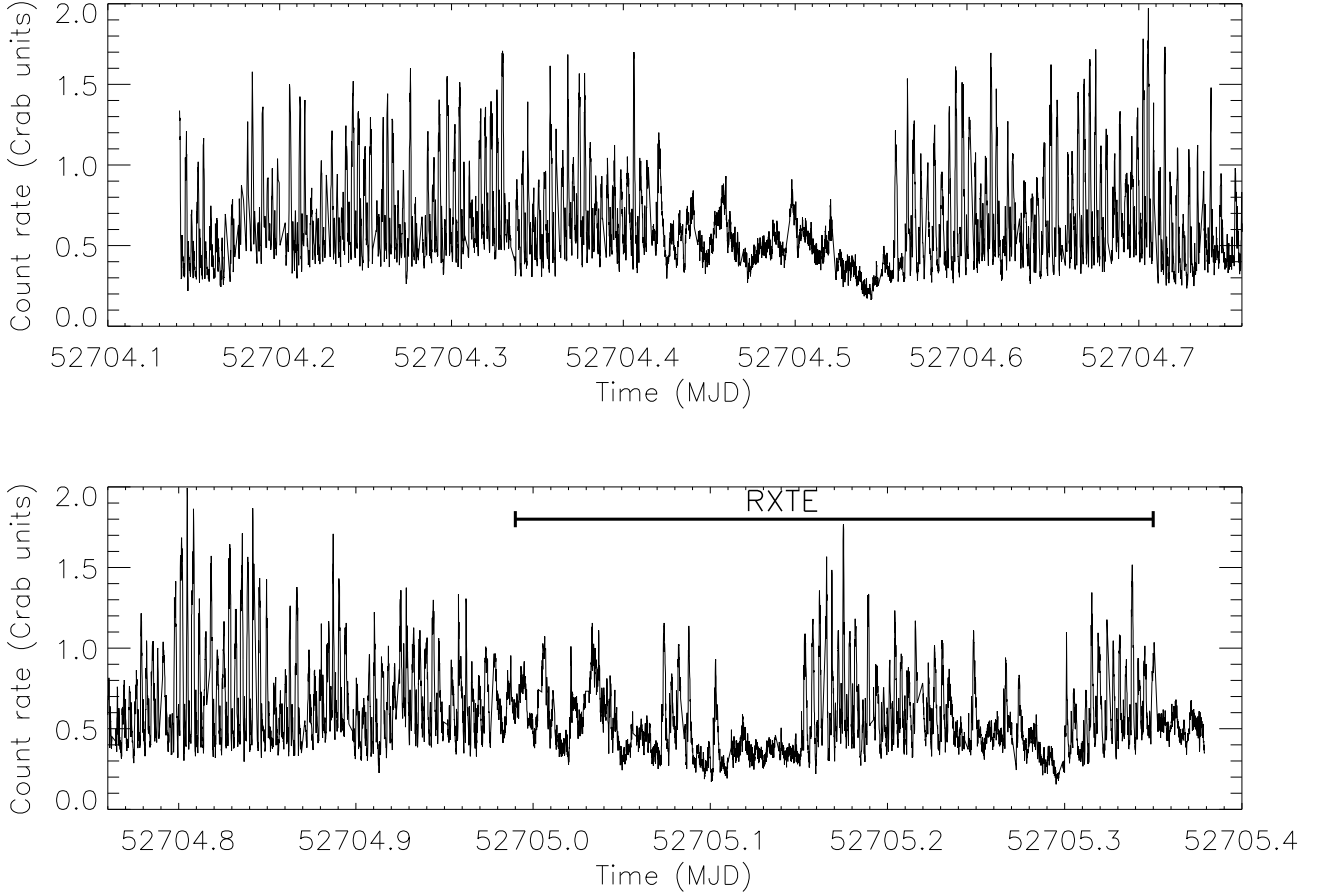}
      \caption{The JEM-X 3--35~keV lightcurve for the whole of
              Revolution 48. The time bin is 8~s. As can be seen,
              the source varied between $\sim$ 0.25--2~Crab. Reproduced from
              Hannikainen et al. (2003). Shown on the plot is our
              simultaneous {\it RXTE} coverage. 
              }
         \label{fig:jemx}
   \end{figure*}

All data were reduced using the Off-line Scientific Analysis
  (OSA) version 4.0 software.
The JEM-X data were reduced following 
  the standard procedure described in the JEM-X cookbook. 
Based on the JEM-X 3--35~keV lightcurve first presented in
   Hannikainen et al. (2003) and reproduced here
   (Fig.~\ref{fig:jemx}), we defined Good Time Intervals (GTIs) 
   for the spectral extraction process. 
We created two GTI files: one for data below 70~counts~s$^{-1}$ 
  (which corresponds to 
  $\sim$ 0.5~Crab in that energy range) and one for data above 70~counts~s$^{-1}$.
We then ran the analysis using the two GTI files from which we
   extracted two total spectra. 
For the modelling of the spectra, we ignored data above 30~keV for the 
  spectrum of the bright parts and above 25~keV 
  for the faint parts.
We assumed systematic uncertainties as shown in Table~\ref{tab:sysjemx}
 (P. Kretschmar \& S. Mart\'{\i}nez N\'u\~nez, priv. comm.).

\begin{table}[htbp]
\centering
\begin{tabular}{ccc}
\hline
\hline
Channel & Energy & Systematic uncertainty\\
         & (keV)  &   (\%)\\
\hline
$<$ 72   & $<$ 5.12 &  10 \\
72--79   &  5.12--5.76 & 5 \\
80--89   &  5.76--6.56 & 2 \\
90--99   &  6.56--7.36 & 7 \\ 
100--109 &  7.36--8.16 & 5 \\ 
110--119 &  8.16--9.12 & 4 \\ 
120--129 &  9.12--10.24 & 5 \\ 
130--149 &  10.24--13.44 & 4 \\ 
150--159 &  13.44--15.04 & 6 \\ 
160--169 &  15.04--17.64 & 5 \\ 
170--179 &  17.64--20.24 & 8 \\ 
180--189 &  20.24--22.84 & 7 \\
190--199 &  22.84--26.24 & 9 \\ 
200--209 &  26.24--29.84 & 15 \\ 
 \hline
\end{tabular}
\caption{Level of systematic uncertainty applied to the spectral
channels of the JEM-X spectra, and energy channel correspondence.}
\label{tab:sysjemx}
\end{table}

\indent  Concerning IBIS, we reduced the data from the top layer
  plane detector, ISGRI (Lebrun et al. 2003).
We made a run of the software up to
  the IMA level, i.e. the production of images. 
During the run we extracted images from individual science windows in
  two energy ranges (20--40~keV and 40--80~keV) as well as a mosaic in
  the same energy ranges. 
As with JEM-X, the GTIs were given as input to OSA to extract the
  ISGRI spectral products. 
Rather than using the standard spectral extraction, we then re-ran
  the analysis software up to the IMA level in ten energy bands and 
  then estimated the spectra.
The source count rate and error were then extracted from each mosaic
  at the position of the source, from the intensity and variance map
  respectively (see the Appendix in Rodriguez et al. 2004b
  for an in-depth description of this method). 
We further assumed 8\% systematics in each channel (e.g. Cadolle-Bel et
  al. 2004). \\
\indent 
To reduce SPI data a pre-release version of OSA 4.0 was used, 
  which incorporates 
  two different analysis methods:  The default pipeline is optimised for 
  speed, but cannot handle user defined GTIs 
  (Kn\"odlseder et al. 2004).
Hence the alternative pipeline, which is more closely 
  based on the OSA 3 release, was adopted but using the more recent 
  calibration data from OSA 4.0.  
The background flux was estimated using the 
  Mean Count Modulation (MCM) method 
  and spectra were extracted from the reduced data, using the same input 
  catalogue as the ISGRI analysis, with SPIROS v8.0.1 (Skinner \&
  McConnell 2003). 

The combined JEM-X, ISGRI and SPI spectra resulted in two broadband
spectra: one for the high part of the JEM-X lightcurve, which we 
  will refer to as pulse maxima and one for the low part, corresponding 
  to the combined pulse minima. 
These two spectra are used for the broadband modelling in Section~\ref{broadspec}.

\subsection{{\it RXTE} data reduction}

\begin{figure}
   \centering
   \includegraphics[width=9cm]{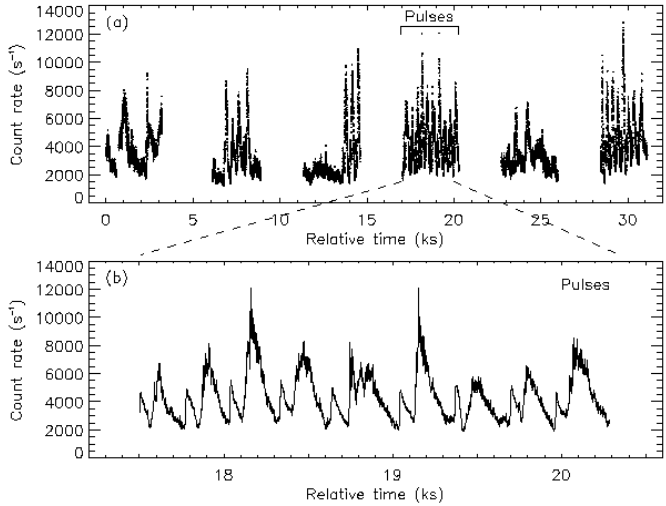} 
      \caption{{\bf a} The {\it RXTE}/PCA 2.5--15~keV lightcurve
	       (with time bin of 1 second) extracted from 
               all PCU and layers.
               The observations start on MJD 52704.99.
               Marked in the plot are the times of the pulses dealt with
               in Figs.~\ref{fig:upsdowns}, \ref{fig:mean9pulse}, 
	       \ref{fig:compare}, \ref{fig:correl}
               and in Section~\ref{rxte}. Each segment in the lightcurve
	       represents one {\it RXTE} orbit.
	       {\bf b} An expanded plot of the pulses.
               }
         \label{fig:pcalite}
   \end{figure}

During the last third of the {\it INTEGRAL} observations we obtained
  simultaneous {\it RXTE} observations ($\sim 32$~ks). 
We reduced and analysed the data from the {\it RXTE} Proportional
  Counter Array using the {\it LHEASOFT} package v5.3,
  following the standard steps explained in the PCA cookbook; 
see Rodriguez, Corbel \& Tomsick (2003) for the details of the selection
  criteria. 
We first extracted a 1-s resolution  lightcurve
  between 2.5--15 keV from the binned  data, from all available
  Proportional Counter Units (PCU) turned on during individual $\sim 
  90$~min {\it RXTE} revolutions. 
The background-corrected lightcurve is shown 
  in Fig. \ref{fig:pcalite}. 
Since we were interested in studying the spectral variations 
  of the source during this observation, we took 
  advantage of the high sensitivity of the PCA to extract spectra on the 
  smallest timescale allowed by the standard 2 data. 
In a manner similar to a previous work (Rodriguez et al. 2002), we 
  extracted PCA spectra and background spectra every 16~s
  from the top layer of PCU 0 and 2. 
The responses were generated with {\sc pcarsp} (V10.1).
The resulting 16-s {\it RXTE} spectra are used to study fast spectral 
  variability in Section~\ref{rxte}.

\section{A new type of variability?}\label{variability}

Fig.~\ref{fig:jemx} shows the entire JEM-X 100~ks lightcurve.
Throughout most of the lightcurve (the high fast flaring parts,
  e.g. from MJD~52704.15--52704.41)
  we see a novel class of variability, characterized 
  by 5-minute pulses, not observed before 
  (Hannikainen et al. 2003).  The entire JEM-X
  lightcurve is dominated by these 5-minute pulses, as evidenced
  by the 3~mHz quasi-periodic oscillation which resulted from a
  Fourier transform the whole JEM-X lightcurve (Hannikainen et
  al. 2003).
Although this kind of pulsed variability resembles the $\rho$-heartbeat
 and $\kappa$ oscillations described in Belloni et al. (2000),
  the oscillations presented there are more uniform and occur on shorter
  timescales. 
 In Hannikainen et al. (2003), we also produced a colour-colour 
   diagram for the 
   {\it RXTE} data in the same manner as in Belloni et
  al. (2000) and showed that the data points fell in a different part
  of the colour-colour diagram as compared to the other classes.
We call this new type of variability $\xi$.

   \begin{figure}
   \centering
   \includegraphics[width=9cm]{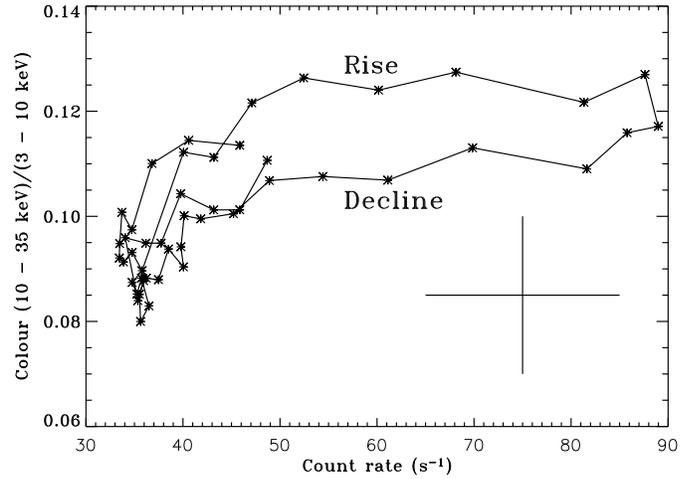} 
      \caption{The JEM-X colour-intensity diagram from the 
              whole 100~ks observation. 
              The colour was constructed
              from the 10--35~keV and 3--10~keV lightcurves. The
              maxima of the flares are
              in the right-hand edge of the plot, while the minima
              are in the left-hand edge. This confirms that the
              rising phases are harder than the declining phases.
              Each point is one time bin (10~s) and 
	      represents the averaged mean value 
	      from the 300 JEM-X flares; the cross shows the
              average standard deviation for the mean value of 
              each individual point
              throughout the loop -- this includes both the
              variability within each point and the error
              (i.e. it is
              not an error bar). 
      }
         \label{fig:upsdowns}
   \end{figure}
  
To study the behaviour of the source in this new variability class,
  we produced a colour-intensity diagram from the whole 100~ks
  JEM-X lightcurve (Fig.~\ref{fig:upsdowns}).
The JEM-X colour was constructed from the 3--10~keV and 10--35~keV
  lightcurves. 
Each point is one time bin of 10~s and represents the 
    averaged mean value from 300 JEM-X pulses.
Points from the pulse-maxima are situated in the right-hand
  side of the plot, while points from the minima are on the left-hand
  side. 
Indicated in the figure are the rising and declining
  phases of the pulses. 
The cross shows the standard deviation of the mean value of 
  each point -- this includes both the real variability and the error,
  and hence is {\it not} an error bar in the conventional sense.
The figure shows not a simple flux-hardness correlation, but rather 
  some kind of limit cycle/hysteretic behaviour with the rising phases 
  being harder than the
  declining phases. This is opposite to what has been observed in
  other similar classes (e.g. class $\rho$). 
In the next section, using the {\it RXTE} data,
  we take a more detailed look at this behaviour,
  starting with an examination of the individual pulses.

\section{Pulses and variability in the {\it RXTE} data}\label{rxte}

\begin{figure}
   \centering
   \includegraphics[width=8cm]{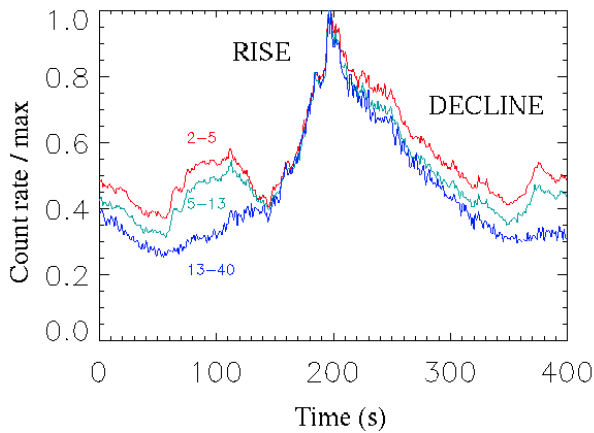}
      \caption{The mean of nine {\it RXTE}/PCA pulses -- the times
              covered by these nine pulses is shown in
              Figs.~\ref{fig:pcalite} and \ref{fig:rxte_spec}. 
              The top (red) line shows the 2--5~keV
              lightcurve, the middle (cyan) the 5--13~keV and the 
	      bottom (blue) the
              13--40~keV lighcurve. The rising phases are shorter and harder
              than the declining phases.  
	      The latter eight pulses were shifted onto the first
              pulse as described in the text, using the maximum of the
              2--5~keV lightcurve as the reference point. 
              }
         \label{fig:mean9pulse}
   \end{figure}

During some of the {\it RXTE} pointings, the 5-minute pulses are clearly
  detected also by {\it RXTE}/PCA (Fig.~\ref{fig:pcalite}a).
In one segment in particular (Fig.~\ref{fig:pcalite}b) we see nine
  consecutive pulses,
  representative of the variability pattern in the majority of the
  JEM-X lightcurve.
Because of the better temporal resolution and higher
  collecting area of {\it RXTE}/PCA compared to {\it INTEGRAL}/JEM-X,
  we can study the
  variability of the source on a timescale of 1~s with good enough statistics.
This time bin is appropriate because Belloni et al. (2000)
  used a time bin of 1~s to make their colour-colour diagrams, hence
  allowing for a direct comparison between this work and theirs.
Another reason why a 1-s binning was chosen is that it is not of too
  high resolution, and therefore does not eliminate too
  many points which would render the plot unclear, but at the same
  time is not so low so as to smear out any variability present.
The {\it RXTE}/PCA lightcurve was divided into three energy bands
  (2--5, 5--13, and 13--40~keV) and smoothed with a boxcar average
  of 5 time bins.
Figure~\ref{fig:mean9pulse} shows the mean of the nine {\it RXTE}/PCA
  pulses from Fig. 3b.
The pulses consist of the main pulse (with the rising
  phase being shorter and harder than the declining phase) and
  a precursor pulse, which is shorter, softer and of smaller
  amplitude than the main pulse.

The times of the maxima of the 2--5~keV pulses were determined from
  the smoothed lightcurve.
The eight latter pulses were shifted onto the first pulse using the
  times of the maxima as the reference.
This was repeated for the other two energy bands, using the times of
  the maxima of the 2--5~keV pulses. 
From the resulting average pulse shape, we can confirm that the rising 
  phase of the (main) pulse is harder than the declining phase.  
In addition, the rise is also shorter than the decline. 
In each of these nine pulses, there is also a precursor seen only in
  the softer X-ray bands. 

  \begin{figure}[t]
   \centering
   \includegraphics[width=9cm]{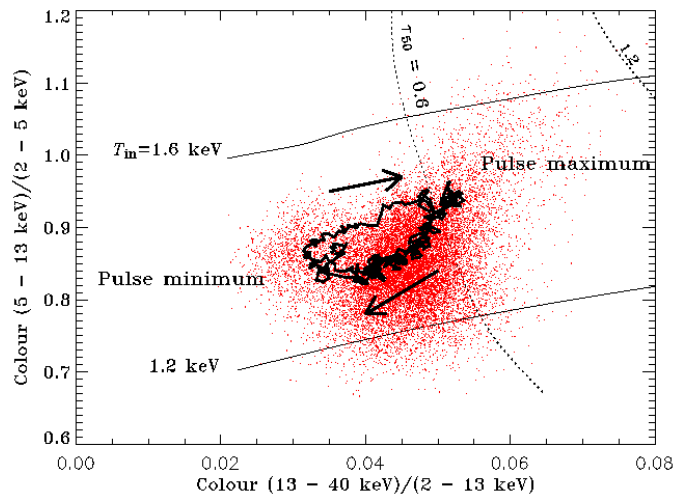} 
        \caption{Colour-colour diagram showing the   
              the {\it RXTE}/PCA from this study.
	      The dots represent the whole observation, while the
              solid line superposed is from the pulses
              (Fig.~\ref{fig:pcalite}b).
	      The curved dotted lines
	      correspond to the colours expected from a Comptonizing
              spherical corona with a given radial optical depth
	      $\tau_{50}$ (i.e. optical depth for the assumed electron
              temperature of 50~keV) and varying values of the inner
              disc temperature $T_{\rm in}$.
	      The solid lines show the colours expected for the model
              for given $T_{\rm in}$ and varying $\tau_{50}$.
              The pulse cycle goes clockwise.
	      }
         \label{fig:compare}
   \end{figure}

To further study the indicated limit cycle, we constructed a 
  colour-colour diagram from the {\it RXTE}/PCA data, shown in 
  Figure~\ref{fig:compare}. 
The X-ray colours were defined as HR1 = B/A and HR2 = C/(A + B), where
  A, B, and C are the counts in the 2--5 keV,
  5--13 keV and 13--40 keV, respectively, following Belloni et
  al. (1997).
In Fig.~\ref{fig:compare}, the dots represent the whole {\it RXTE} 
  observation (including the non-pulsed part), while the solid line 
  shows the cycle traced by the pulses (Fig.~\ref{fig:pcalite}b).
The main aim of the figure is to show the small range of variability of the
  5-minute pulses.
The pulse cycle goes clockwise as indicated by the arrows.
The maxima of the pulses are at the upper-right of the pulse cycle
  curve, while the minima are at the lower left.

Superposed in Fig.~\ref{fig:compare} are model curves from a study by
  Vilhu \& Nevalainen (1998), calculated as explained below.
Although a direct comparison of the parameters is not possible  
  due to the {\it RXTE}/PCA
  gain evolution through the years, resulting in a change of the energy channel
  correspondence, we simply want to illustrate qualitatively the
  position in
  the colour-colour diagram of this present observation. 
In order to have an idea of what physical parameters the 
  observed colours correspond to, we simulated the colours predicted 
  by a Comptonization model.
We assumed the Comptonizing source to be a sphere surrounded by 
  a cooler disc with the inner radius equal to that of the sphere 
  (Poutanen, Krolik, \& Ryde 1997).
Seed photons coming from the disc are characterized by the inner disc
  temperature $T_{\rm in}$.
The electrons in the cloud were assumed to be thermal with temperature
  $kT=50$ keV. 
The Thomson optical depth of the cloud along the radius is $\tau_{50}$,
  with subscript 50 meaning that this $\tau$ corresponds to the 
  50 keV electrons.
For a different $T$, similar Comptonization spectra are produced
  by a cloud of optical depth $\tau=\tau_{50}\times 50 {\rm keV}/kT$,
  because the slope of the Comptonization spectrum, $\Gamma$, is a 
  function of the Kompaneets parameter, $y=4\tau kT/m_e c^2$, and
  $\Gamma=\frac{9}{4}y^{-2/9}$ (Beloborodov 1999).
For the given $T_{\rm in}$ and $\tau_{50}$, we compute the 
  Comptonization spectrum using the method of Poutanen \& Svensson
  (1996). 
Then we obtain the colours expected from
  the model spectrum using the responses of {\it RXTE}/PCA and plot
  them on the colour-colour diagram (Fig.~\ref{fig:compare}). 
We see that the pulses discussed in this paper correspond to 
  optical depth $\tau\la 0.7$ and inner disc temperature 
  of $T_{\rm in}\sim 1.3 -  1.4$ keV.
This temperature differs slightly from that obtained using a black body $+$
  power law model, 
  because the Comptonization spectrum is not a
  power law close to the seed photon energies (see e.g. Vilhu et
  al. 2001). \\

\subsection{Hysteresis or limit-cycle oscillations?}

  \begin{figure*}
   \centering
   \includegraphics{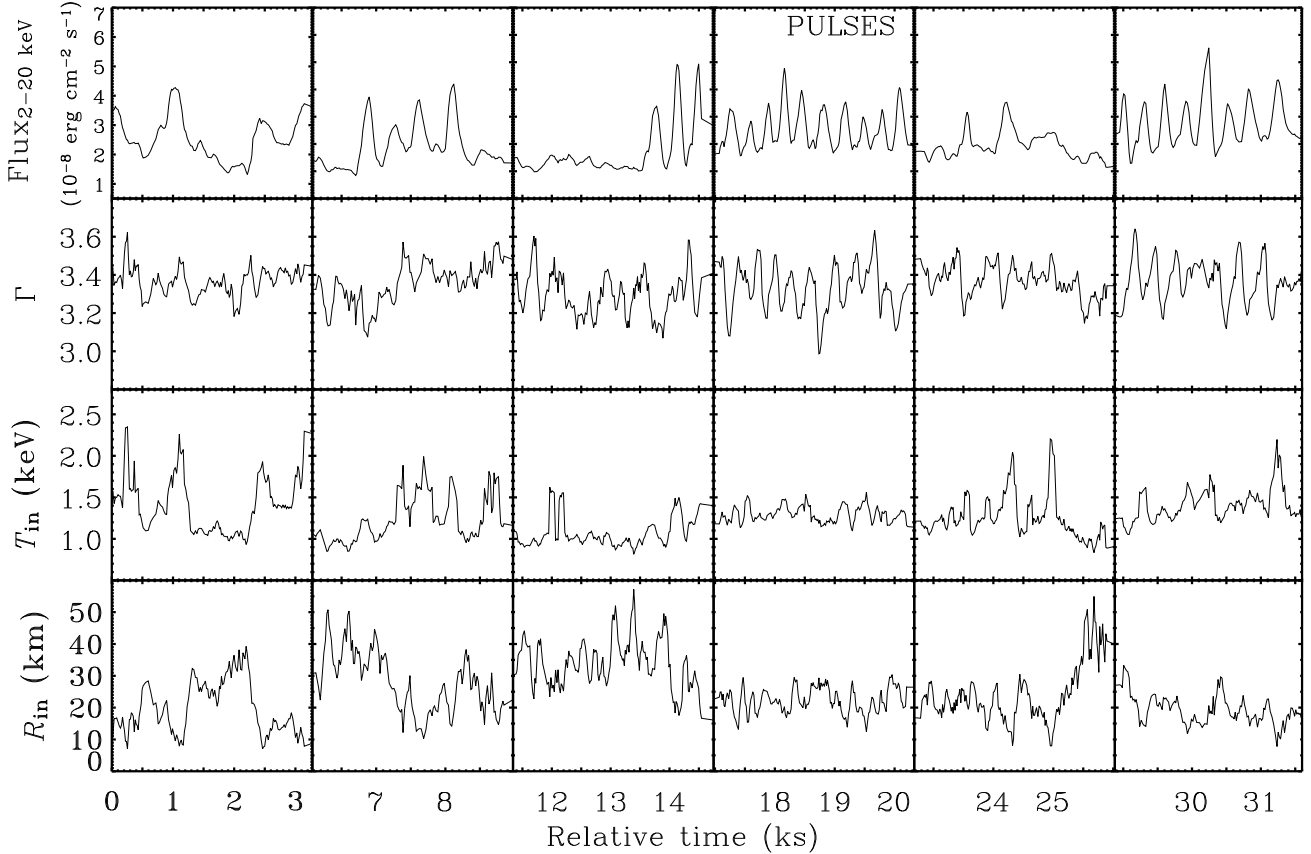}
      \caption{The results from the {\it RXTE} spectral analysis. 
	Each column represents one {\it RXTE} orbit -- the fourth
	column from the left is from the time of the pulses examined in 
      more detail. The
      top row reproduces the PCA 2--20~keV flux. The second row
      from the top shows the evolution of the photon index $\Gamma$ 
      throughout the observations, the third row shows the inner disc
      temperature (keV), while the bottom row shows the inner disc
      radius (km). All results are from the fits to the 16-s spectra. 
      Bad fits, i.e. data points with $\chi^2>50$ (for $\sim 41$
      d.o.f.), were removed.  
      The remaining data and fit results were smoothed with a boxcar 
      average of 5 time bins to show the general trend throughout the data
      sets. 
              }
         \label{fig:rxte_spec}
   \end{figure*}

\begin{figure*}
\centering
\includegraphics[width=16cm]{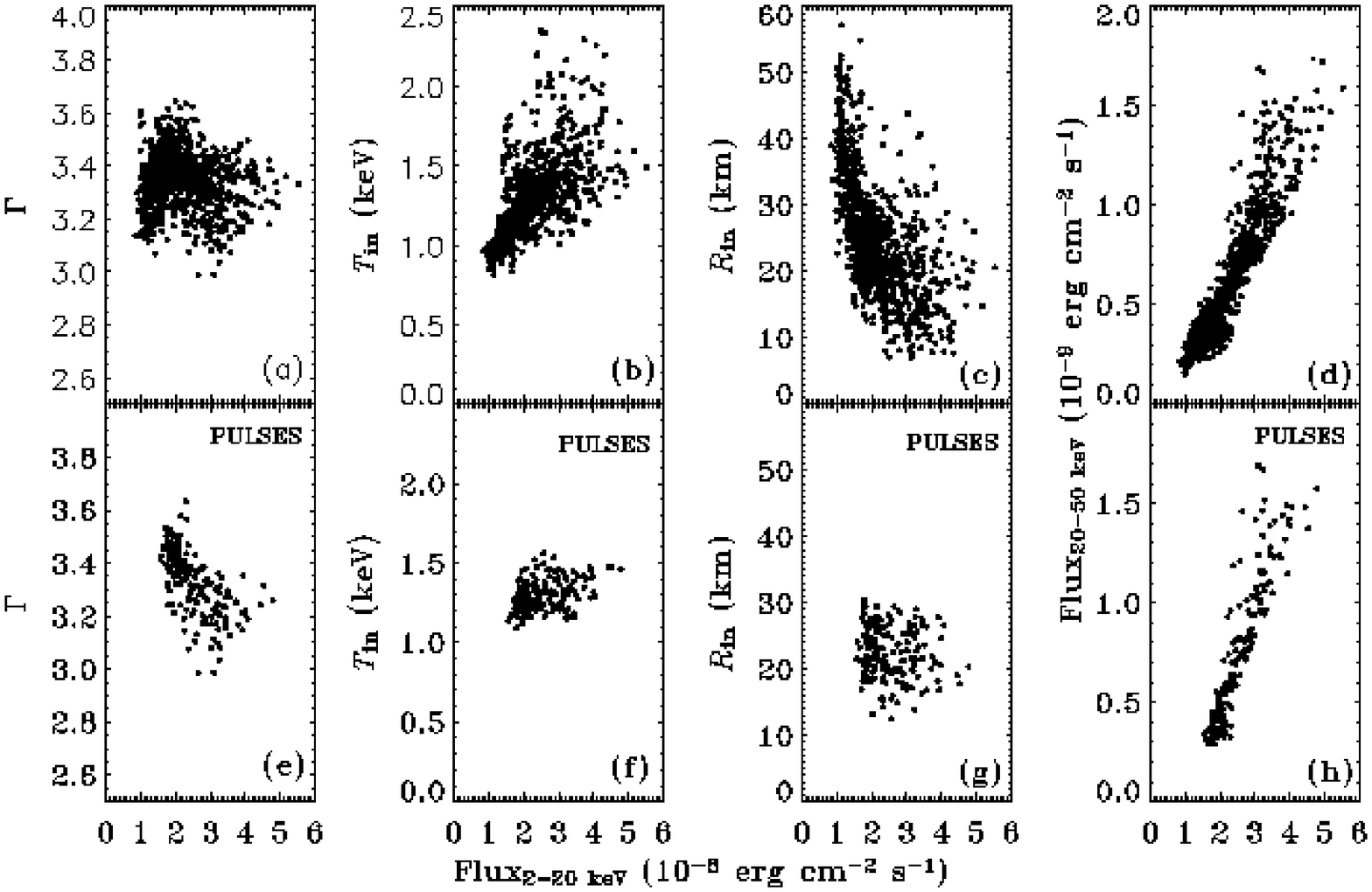}
   \caption{Panels {\bf a --d} show $\Gamma$, $T_{\rm in}$, $R_{\rm
           in}$ and the 20--50~keV flux 
           against the 2--20~keV flux from the whole set of the
           {\it RXTE}/PCA observations, while panels {\bf e--h} show the
           same but only for the pulses.
           }
   \label{fig:correl}
\end{figure*}

The colour-intensity diagram in Fig.~\ref{fig:upsdowns} and 
  the colour-colour diagram in Fig.~\ref{fig:compare} both show 
  loop-like features, 
  which could be due to a hysteretic process or limit-cycle oscillations.  
The hysteresis phenomenon has been suggested 
  to explain the loop features in the X-ray colour-colour diagram  
  of black hole low-mass X-ray binaries (Miyamoto et al. 1995).
One recent example is GX 339$-$4 (see Zdziarski et al.\ 2004).  
Hysteresis can occur on a large range of timescales, 
  and for accreting black holes the timescale is attributed 
  to the finite time required in the building up and dissipation 
  of a viscous, optically thick accretion disc.  
The accretion disc gives rise to low-energy photons, 
  which act as coolants for the hot plasma via inverse Comptonization. 
During the accretion disc build-up process, 
  the source luminosity increases but the X-ray spectrum softens. 
On the other hand, when the accretion disc dissipates 
  (e.g. due to evaporation),  
  the source luminosity decreases and the X-ray spectrum hardens.  
The process is expected to occur in an extended region 
  at a substantial distance from the central black hole.  
The duration of the two hysteretic states and their transition 
  times are long, 
  and for low-mass X-ray binaries the timescale would be 
  a year or more. 

In \grs1915, the timescale for developing the loop features 
  is relatively short -- roughly hundreds of seconds. 
The observation is not consistent with the scenario 
  requiring an extended accretion disc, 
  the building up and the recession of which drive the hysteretic cycle. 
One may argue that the process could occur in a smaller inner disc region,   
  and this naturally shortens the corresponding timescales.  
When we examined the loop in Fig.~\ref{fig:upsdowns}, 
  we find spectral hardening accompanied by an increase in the X-ray
  photon counts, 
  contrary to the properties of the hysteresis model 
  for the GX 339$-$4 behaviour. 
We therefore conclude that the processes behind the loops 
  in the colour-colour diagrams of \grs1915 and GX 339$-$4 are not identical. 
As an alternative, we interpret the loops in the \grs1915 data 
  as limit-cycle oscillations, 
  which are probably due to delayed feedback in the inner accretion disc.  
A possible explanation for the simultaneous rise in the photon counts 
  and spectral hardening 
  is an increase in the size of the hard X-ray emission region  
  which produces the Comptonized hard X-rays. 
This region could be a hot disc corona 
  or some non-thermal plasmas in the vicinity of the accreting black hole.   

\subsection{Fast spectral variability}\label{rxte_spectra}

To track variability on short timescales, the 16-s PCA spectra were fitted 
  individually in {\sc XSPEC} V11.3 (Arnaud 1996)  
  with a simple model of absorbed power law plus multicolour disc 
  blackbody (MCD, Mitsuda et al. 1984), and a Gaussian.
The absorption column density $N_{\rm H}$ was fixed at $5\times 
  10^{22}$~cm$^{-2}$. 
The power law photon index $\Gamma$, inner disc 
  temperature $kT_{\rm in}$ and inner disc radius $R_{\rm in}$ are plotted in
  Fig. \ref{fig:rxte_spec}, in addition to the actual lightcurve in
  flux units.
All bad fits (points with $\chi^2>50$ for $\sim 41$ d.o.f.) were
  removed and the data were smoothed with a boxcar average of 5 time bins to
  highlight the general trend in the data sets.
We clearly see amplitude variations of 
  the fit parameters related to variations in the 2--20~keV 
  flux.
The inner disc radius is obtained from the normalisation, $N$, of the
  MCD model following 
  $R_{\rm in}=d_{10} \times \sqrt{N/\cos\theta}$, where 
  $\theta$ is the inclination angle, and $d_{10}$ is the distance in 
  units of 10 kpc.
For plotting purposes, we used 
  $\theta=70^\circ$ and $d_{10}=1$ (Chapuis \& Corbel 2004 quote a
  distance range of 6--12~kpc).
No colour correction (e.g. Shimura \& Takahara 1995) has
  been applied to those values. 
We remark that sometimes the inner disc 
  temperature shows high values ($kT_{\rm in}>1.5$ keV), 
  along with very low values of the inner disc radius. 
Those very high $kT_{\rm in}$ spikes are indicative of times when the
  MCD model is inappropriate (Merloni, Fabian \& Ross 2000). However, 
  the simple model we used in our 
  fitting allows us to track the general spectral variations of the source on 
  these short timescales, and even if the MCD parameters are unphysical during
  the spikes, they are indicative of underlying changes in the source spectrum.

Fig.~\ref{fig:correl} shows correlation plots between $\Gamma$, 
  $T_{\rm in}$, $R_{\rm in}$ and the 20--50~keV flux against
  the 2--20~keV flux for the whole set of {\it RXTE}/PCA observations
  (a--d) and only for the subset of pulses (e--h) fitted with the
  simple MCD$+$power law model.
The overall {\it RXTE}/PCA data sets (panels a--d) 
  represent a mixture of source behaviour 
  (see the lightcurve in Fig.~\ref{fig:pcalite} for
  example), as the {\it RXTE}/PCA observations covered times when the
  source behaviour did not necessarily exhibit the new variability
  pattern.
Fig.~\ref{fig:correl}a shows that there is no obvious correlation
  between the photon index and the 2--20~keV flux throughout the
  whole observation.
Panel b shows that there is the tendency for $T_{\rm in}$ to increase
  with increasing flux, while panel c 
  shows that $R_{\rm in}$ decreases with increasing
  flux. 
The data sets from the pulses are slightly different from the overall
  data set which includes the non-pulsed part. 
To begin with, Fig.~\ref{fig:correl}e shows that on the whole the
  photon index decreases with increasing flux, which supports the
  scenario presented, for example, in Fig.~\ref{fig:mean9pulse} 
  that the maxima are harder than the minima.
Fig.~\ref{fig:correl}f shows that there is no substantial variation in $T_{\rm
  in}$ with flux, indicating that the pulses are largely isothermal.
Again, Fig.~\ref{fig:correl}g shows no obvious correlation between
  $R_{\rm in}$ and flux, i.e.
  $R_{\rm in}$ does vary, but not coherently with the flux.

On the other hand, Fig.~\ref{fig:correl}d and h show a very strong
  linear correlation between the flux in the 2--20~keV range and the
  20--50~keV range, both for the overall data set and for the subset of
  pulses. 
Taken together with the photon index--flux anticorrelations, this
  implies that there is no pivoting in the {\it RXTE} energy range,
  but the flux at higher energies increases more than at lower
  energies.

\section{Broadband spectral modelling}\label{broadspec}
\subsection{Spectral Model}

In order to conduct a more detailed analysis of the spectrum, we used 
the co-added spectra, extracted from the high and low flux parts in the {\it INTEGRAL}
  lightcurve as described in Section~\ref{datared}. 
This provides us with two broadband 
  spectra (JEM-X $+$ ISGRI $+$ SPI) in the energy range of 3--200 keV, 
  for the pulse maxima and minima separately. 
The two spectra were fitted using {\sc XSPEC} V11.3 (Arnaud 1996). 
We interpret the spectra in terms of Comptonization of soft X-ray seed 
photons, assumed here to be a disc blackbody with a maximum 
temperature, $T_{\rm in}$. We use a Comptonization model by Coppi 
(1992, 1999), {\tt eqpair}. 
This model has previously been used to fit X-ray spectra of \grs1915 
(based on \pca/HEXTE and \gro/OSSE data) by Zdziarski et al.\ (2001), 
Cyg~X-1 (Poutanen \& Coppi 1998;  Gierli\'nski et al.\ 1999),
as well as \integral\/ and \xte\/ data of Cygnus X-3 
(Vilhu et al.\ 2003).

\begin{table*}
      \caption{Model parameters$^{\mathrm{a}}$ for the best-fit 
models to the broadband spectra for the co-added pulse maxima and minima.}
         \label{parameters}
     $$   
         \begin{array}{lcccccccccccc}
            \hline \hline
            \noalign{\smallskip}
{\rm Data} & N_{\rm H} & kT_{\rm in} & \lh/\ls  & \lnth/\lth  & {\rm
            \Gamma_{inj}} & \tau_{\rm i} & \tau^{\mathrm{b}} & kT_{\rm
            e}^{\mathrm{b}} & \Omega/ 2\pi^{\mathrm{c}} &
 F_{\rm K \alpha} & {F_{\rm bol}}^{\mathrm{d}} & \chi^2/\nu \\
            \noalign{\smallskip}
& 10^{22}\, {\rm cm}^{-2} & {\rm keV} &&&&&& {\rm keV} && {\rm cm}^{-2}\, {\rm s}^{-1} & {\rm erg\,
            cm}^{-2}\, {\rm s}^{-1}&\\
            \noalign{\smallskip}
            \hline
            \noalign{\smallskip}
{{\rm Pulse~maxima}} & 5.4_{-0.9}^{+0.7} & 1.07^{+0.10}_{-0.40} & 0.21_{-0.02}^{+0.03} &
            0.26_{-0.07}^{+0.02} & 1.0_{-0.9}^{+0.2} &  0^{+0.63}
            & 0.50 & 28.5 & 1.5_{-0.15}^{+0.15} &
            2.2\times 10^{-3} & 5.8\times 10^{-8} & 87/191\\
             \noalign{\smallskip}
{{\rm Pulse~minima}} & 6.4_{-0.9}^{+1.4} & 1.05^{+0.08}_{-0.24} & 0.12_{-0.01}^{+0.03} &
            0.21_{+0.04}^{-0.11} & 1.9_{-0.5}^{+0.6} &  0.18_{-0.15}^{+0.20} &
            0.34 & 34.4 & 2.0_{-1.0} & 
            1.3\times 10^{-2} & 2.9\times 10^{-8} & 93/176\\
            \noalign{\smallskip}
            \hline
         \end{array}
     $$ 
\begin{list}{}{}
\item[$^{\mathrm{a}}$] The uncertainties are for 90\% confidence, i.e., $\Delta \chi^2=2.71$.
\item[$^{\mathrm{b}}$] Calculated from the energy and pair balance, i.e., not a free fit parameter. 
\item[$^{\mathrm{c}}$] Assumed $\leq 2$ in the fits.
\item[$^{\mathrm{d}}$] The bolometric flux of the unabsorbed model spectrum.
\end{list}
\end{table*}

The electron distribution in the {\tt eqpair} model can be purely 
 thermal or hybrid, i.e., 
  Maxwellian at low energies and non-thermal at high energies, if 
an acceleration 
process is present. The electron distribution, including the
temperature, $T_{\rm e}$, 
is calculated self-consistently from the assumed form of the acceleration (if 
present) and from the luminosities corresponding to the plasma heating rate, 
$L_{\rm h}$, and to the seed photons irradiating the cloud, $L_{\rm
  s}$.  
The 
total plasma optical depth, $\tau$, includes contributions from
electrons formed 
by ionization of the atoms in the plasma, $\tau_{\rm i}$ (which is a free 
parameter) and a contribution from e$^\pm$ pairs, $\tau-\tau_{\rm i}$ 
(which is 
calculated self-consistently by the model). The importance of pairs and the 
relative importance of Compton and Coulomb scattering depend on the
ratio of the 
luminosity, $\ell$, to the characteristic size, $r$, which is usually expressed in 
dimensionless form as the compactness parameter, 
$\ell \equiv L\sigma_{\rm T}/(r 
m_{\rm e} c^3)$, where $\sigma_{\rm T}$ is the Thomson cross 
section and $m_{\rm 
e}$ is the electron mass. The compactnesses corresponding  to the electron 
acceleration at a power law rate with an index, $\Gamma_{\rm inj}$ and to a 
direct heating (i.e., in addition to Coulomb energy exchange with non-thermal 
\ee\ and Compton heating) of the  thermal \ee\ are denoted as $\lnth$ and 
$\lth$, respectively, and $\lh=\lnth +  \lth$. 
Details of the model are given in 
Gierli\'nski et al.\ (1999).

Following Zdziarski et al.\ (2001), we assume here a constant
$\ls=100$, 
compatible with the high luminosity of \source. For example, for 1/2
of the 
Eddington luminosity, $L_{\rm E}$, and spherical geometry, the size of
the 
plasma corresponds then to $r\sim 100GM/c^2$. We note, however, that
the 
dependence of the fit on $\ls$ is rather weak, as this parameter is 
important only for \ee\ pair production and Coulomb scattering, with 
the former not constrained by our data and the latter important only 
at $\ls\la 1$ or so (Gierli\'nski et al.\ 1999). 

We include Compton reflection (Magdziarz \& Zdziarski 1995),
parametrized 
by an 
effective solid angle subtended by the reflector as seen from the hot plasma, 
$\Omega$, and an Fe K$\alpha$ fluorescent line from an accretion disc 
assumed to 
extend down to $6 GM/c^2$. We assume a temperature of the reflecting 
medium of $10^6$ K, and allow it to be ionized, using the ionization 
calculations of Done et al.\ (1992). We define the ionizing parameter 
as $\xi_{\rm ion}\equiv 4 \pi F_{\rm ion}/n$, where $F_{\rm ion}$ is the
ionizing 
flux and $n$ is the reflector density. 
As the data poorly constrain $\xi_{\rm ion}$, 
  but clearly require the reflector to be ionized, 
  we freeze it to a value of 100 erg~cm~s$^{-1}$, 
  in the middle of the confidence intervals for both fits.

We further assume the elemental abundances of Anders \& Ebihara (1982), an 
absorbing column of $N_{\rm H} \geq 1.8 \times 10^{22}$~cm$^{-2}$, 
  which is an estimate of 
the Galactic column density in the direction of the source by 
Dickey \& Lockman 
(1990), and an inclination of $66^\circ$ (Fender et al.\ 1999).

\subsection{Results}

\begin{figure}
   \centering
   \includegraphics[width=8cm]{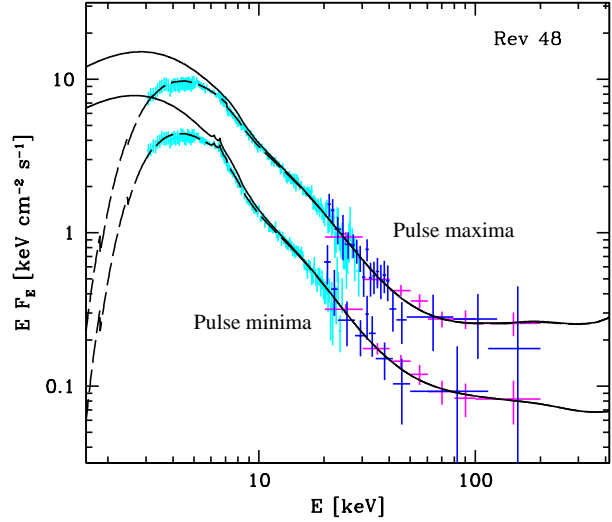}
   \caption{Deconvolved spectra of GRS 1915+105 during Revolution 48
   from the pulse maxima (top) and minima (bottom) separately. 
   The cyan, magenta and blue
   datapoints are from the JEM-X, ISGRI and SPI, respectively. The
   ISGRI and SPI data were renormalized to the JEM-X. The dashed and
   solid curves represent the best-fit models to the observed and
   unabsorbed spectra respectively, with the parameters given in Table 
  \ref{parameters}. }
\label{spectra}
    \end{figure}

\begin{figure}
   \centering
   \includegraphics[width=8cm]{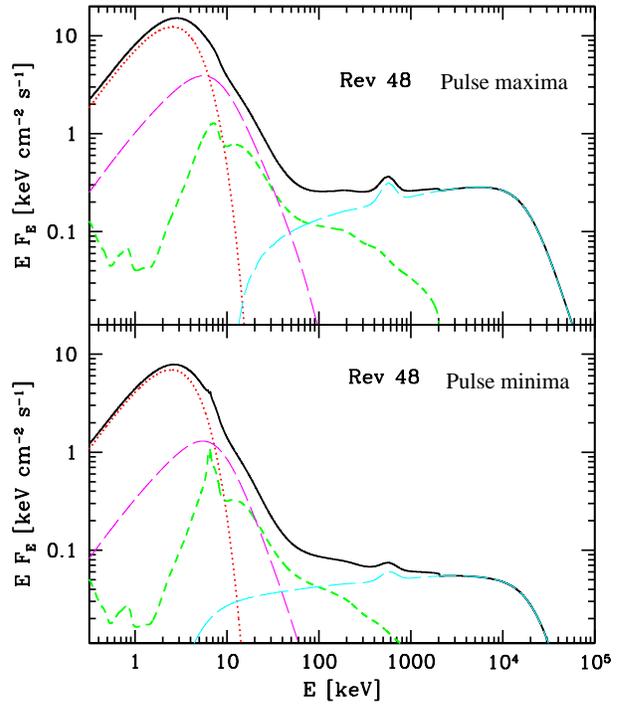}
   \caption{Spectral components of the fits to the pulse maxima (upper panel)
   and minima (lower panel). The dotted and long-dashed curves
   show the unscattered blackbody, and Compton scattering from thermal
   (red) and non-thermal electrons (cyan), including a component from \ee\
   pair annihilation, important around 511 keV. The short-dashed
   (green) curve shows the component from Compton
   reflection, including the Fe K${\alpha}$ line.}
\label{components}
    \end{figure}

The two spectra are shown in Fig.\ \ref{spectra} together with the best-fit 
model. The parameters for the best-fit models are shown in Table
\ref{parameters}, and Fig.\ \ref{components} shows the spectral
components of the fits for the two spectra. 
Note that the small values for $\chi^2$ are due to the systematics
  assumed for the {\it INTEGRAL} data.

Figure\ \ref{spectra} shows that the main difference between the pulse maxima and 
minima is in the observed flux, which is higher by a factor of
$\simeq$2 during maxima. Both spectra show that \source\ during Revolution 48 was in a soft
state, with spectra similar to that of state A in Sobolewska \& \.Zycki (2003)
and in between the softest spectrum in the variability class C/$\chi$ and
that in the B/$\gamma$ class, as observed by \xte\/ and \gro\/ in 
Zdziarski et al. (2001).

The fits imply (Fig.\ \ref{components}) that during pulse maxima as well as minima the 
  spectrum at $\la$6 keV is dominated by the unscattered disc emission, with 
  Comptonized emission dominating only at higher energies, and including a 
  significant contribution from non-thermal electrons. 
The figure shows scattering by the thermal and the non-thermal
  electrons separately.  
This is obtained by first calculating the scattering spectrum by the
  total hybrid distribution, consisting of a Maxwellian and a
  non-thermal high-energy tail. 
Then, the non-thermal compactness is set to zero, $\ell_{\rm nth}=0$, 
  leaving us with only thermal plasma. 
However, since the plasma parameters, $\tau$ and $kT_{\rm e}$, are determined 
  by the model self-consistently, they change after setting 
  $\ell_{\rm nth}=0$. 
Thus we have to adjust the model input parameters so as to 
  recover the original $\tau$ and $kT_{\rm e}$. 
Since there is substantial pair production in the hybrid model 
  but virtually none in the thermal model, we can set the (input) 
  parameter $\tau_{\rm i}$ equal to the (output) $\tau$ of the 
  hybrid model. 
We then need to adjust the plasma cooling rate, determined by 
  $\ell_{\rm h}/\ell_{\rm s}$. 
Thus, we adjust that ratio to a value at which $kT_{\rm e}$ is equal 
  to the original value of the hybrid model. 
At this point, we calculate the thermal scattering spectrum, and subtract 
  it from the total scattering spectrum to obtain the non-thermal component. 
A Compton reflection component including a strong Fe K$\alpha$ line 
  is also required by the fits to both spectra. 

We see, however, that there is significant softening of the 
Comptonized part of the spectrum during pulse minima. This is reflected in the 
ratio, $\lh/\ls$, being significantly smaller. The ratio 
$\lh/\ls$, or, equivalently, $L_{\rm h}/L_{\rm s}$, is between the 
power supplied to
the electrons in the Comptonizing plasma and that in soft disc
blackbody photons irradiating the plasma. Our fits do not require an
additional component in soft photons not irradiating the hot plasma
and thus $L_{\rm s}$ corresponds to the luminosity in the entire disc 
blackbody emission.

As implied by the value of the bolometric flux in Table \ref{parameters}
(see also Fig.\ \ref{spectra}), the value of $L_{\rm s}$ (which
dominates the bolometric flux) decreases by a factor of $\simeq 2$ during minima. 
Given the corresponding decrease of $L_{\rm h}/L_{\rm s}$ by a 
factor of $\simeq 2$, we find that $L_{\rm h}$ decreased by $\sim 4$. 
Conversely, there is an 
increase in the relative power supplied to the electrons in the plasma 
during pulse maxima by a factor of $\sim$2. 

It should be noted that the spectrum from the pulse maxima contains both the end of the
rising phase and the beginning of the declining phase of each
pulse, and thus it is an average of the harder rising and the softer declining
phases. 

The other best-fit parameters for the two spectra are consistent with 
being the same within the measurement uncertainties. 
Given the importance of \ee\ pair production, 
the contribution to the total optical depth, $\tau$, from the
ionization 
electrons is relatively weakly constrained, especially in the
high-flux state, when it is consistent with being null (i.e., with all
scattering plasma formed by pair production). The corresponding range 
of the total optical depth, $\tau$, is much narrower; 
however, the {\tt eqpair} model does not allow 
the uncertainties in $\tau$ to be calculated directly as it
is not a free parameter.

\section{Summary}\label{concl}

We have presented {\it INTEGRAL} and {\it RXTE} data for \grs1915 from
  {\it INTEGRAL's} Revolution 48 in 2003 March.
The source was found to exhibit a novel type of variability 
   which we call $\xi$.
The continuous 100 ks observations, due to {\it INTEGRAL's}
  eccentric orbit, allowed us to classify this new
  variability that dominated the entire JEM-X 3--35 keV lightcurve,
  and is characterized by repeated 5-minute pulses. 
The simultaneous {\it RXTE}/PCA observations showed in one pointing 
  a series of nine $\sim 5$~minute
  pulses, corresponding to the pattern seen by JEM-X. 
By examining the source behaviour in terms of flux-hardness 
  correlations as well as individual pulse shape, we have shown 
  that the rising phases of the pulses are harder and shorter than the
  declining phases, contrary to what is observed in the $\rho$ and 
  $\kappa$ classes of
  Belloni et al. (2000) which this new variability class otherwise resembles. 
Simple modelling of the fast spectral variability confirms a negative 
  correlation between hardness and flux but shows no obvious 
  correlation between flux and inner disc temperature or radius, 
  suggesting that the pulses are largely isothermal. 
Loop-like features, traced out by the pulses in the colour-colour 
  diagram are interpreted as limit-cycle oscillations probably due 
  to delayed feedback in the inner accretion disc.
 
Broadband spectral modelling of co-added spectra for the pulse 
  maxima and minima respectively, using {\tt eqpair}, showed that 
  the spectra of maxima and minima both belong to spectral state 
  A, according to the classification by Belloni et al. (2000), and 
  were essentially similar,
  except for a significant softening of the
  Comptonized part of the spectrum during minima.
The similarity of the spectra was also confirmed in the plot of the
  2--20~keV against the 20--50~keV PCA fluxes which was essentially
  linear, both throughout the whole observation and solely during the
  pulses. 

We have also seen that there is essentially no correlation
  between $T_{\rm in}$ and the flux during the pulses, 
  again supporting the suggestion that they are isothermal. 
This implies that this new variability pattern in itself is largely
  isothermal by nature. 
The fact that the spectra do not change between the pulse maxima and the pulse
  minima (except for the flux level) and that the pulses are described
  by a fast rise and a slow decline indicates that the variability is
  not representative of oscillations between spectral states (as
  e.g. class $\rho$ in Belloni et al. (2000)).
We attributed the variability to limit-cycle oscillations
  which are probably due to delayed feedback in the inner accretion
  disc.
However, the variability could also result from flares in the
  accretion disc arising from instabilities in the accretion rate.
  
 Recently, Tagger et al. (2004) have shown, based on the work by
  Fitzgibbon et al. (priv. comm.), that \grs1915 always seems to follow the
  same pattern of transition through all 12 classes as defined by
  Belloni et al. (2000).
This may suggest that \grs1915 simply displays a repeating continuum of all
  possible configurations and that we caught it in an intermediate
  class between two previously known ones.  
This implies that more variability classes remain to be observed 
  and that, with continued monitoring of \grs1915,
  we should be able to observe other classes of variability 
  behaviour and gain a better understanding of this 
  mystifying source.

\begin{acknowledgements}
DCH thanks Sami Maisala for installing pre-OSA4.0
  at the Observatory, University of Helsinki,
  and Juho Schultz for help with programming.
JR thanks Marion Cadolle-Bel, Andrea Goldwurm and Aleksandra Gros 
  for useful discussions and help with the IBIS analysis. 
We thank the anonymous referee for useful comments.
DCH, LH and JP acknowledge support from the Academy of Finland.
JR acknowledges financial support from the French Space Agency (CNES).
AAZ has been supported by KBN grants PBZ-KBN-054/P03/2001, 1P03D01827 and
  by the Academy of Finland.
This work was supported in part by the NORDITA Nordic
  project on High Energy Astrophysics and by the HESA/ANTARES programme
  of the Finnish Academy and TEKES, the Finnish Technology Agency.
Based on observations with {\it INTEGRAL}, an ESA project with 
  instruments and science data center funded by ESA and member states 
  (especially the PI countries: Denmark, France, Germany, Italy,
  Switzerland, and Spain), the Czech Republic, and Poland and with the 
  participation of Russia and the USA.
We acknowledge quick-look results from the {\it RXTE}/ASM team. 
This research has made use of NASA's Astrophysics Data System.
\end{acknowledgements}


\begin{thebibliography}{}

\bibitem{ae82}
Anders, E., \& Ebihara, M. 1982, Geochim.\ Cosmochim.\ Acta, 46, 2363

\bibitem{a96}
Arnaud, K. A. 1996, in Jacoby G. H., Barnes J., eds., Astronomical Data
Analysis Software and Systems V, ASP Conf. Series Vol.\ 101, San Francisco,
p.\ 17

  \bibitem[Belloni et al. 1997]{belloni97}
    Belloni, T., M\'endez, M., King, A.R., van der Klis, M.,
       \& van Paradijs, J. 1997, ApJ, 488, L109

    \bibitem[Belloni et al. (2000)]{belloni}
      Belloni, T., Klein-Wolt, M., M\'endez, M., van der Klis, M.,
       \& van Paradijs, J. 2000, A\&A, 355, 271

\bibitem[Beloborodov]{belo}
 Beloborodov, A.M. 1999, in ASP Conf. Ser. 161, High Energy 
  Processes in Accreting Black Holes, ed. J. Poutanen \& R. Svensson 
  (San Francisco: ASP), 295

\bibitem[Cadolle-Bel et al. 2004]{cadolle-bel}
       Cadolle-Bel, M., Rodriguez, J., Sizun, P., et al. 2004,
       A\&A, 426, 659

   \bibitem[Castro-Tirado et al.(1992)]{castro}
       Castro-Tirado, A.J., Brandt, S., \& Lund, N.
                          1992, IAUC 5590

\bibitem[Chapuis \& Corbel (2004)]{chapuis}
       Chapuis, C., \& Corbel, S. 2004, A\&A, 414, 659

\bibitem{c92}
Coppi, P.S. 1992, MNRAS, 258, 657

\bibitem{c99}
Coppi, P.S. 1999, in ASP Conf.\ Ser.\ Vol.\ 161, High
Energy Processes in Accreting Black Holes, ed.\ J. Poutanen \& R. Svensson
(San Francisco: ASP), 375

\bibitem[Courvoisier et al. (2003)]{courvoisier} 
     Courvoisier, T. J. L., Walter, R., Beckmann, V., et al. 2003, 
     A\&A, 411, L53

\bibitem[Dhawan et al. (2000)]{dhawan}
     Dhawan, V., Mirabel, I.F., \& Rodr\'\i guez, L.F. 2000, ApJ, 543, 373

\bibitem[Dickey \& Lockman 1990]{dickey90}
   Dickey, J.M., \& Lockman, F.J. 1990, ARA\&A, 28, 215

\bibitem{d92}
Done, C., Mulchaey, J. S., Mushotzky, R. F., \& Arnaud, K. A. 1992, \apj, 395,
275

  \bibitem[Fender et al. (1999)]{fender99} Fender, R.P., Garrington,
   S.T., McKay, D.J., et al. 1999, MNRAS, 304, 865

\bibitem[Fender \& Belloni (2004)]{fender04}
   Fender, R., \& Belloni, T. 2004, ARA\&A, 42, 317

  \bibitem[Fuchs et al. (2003)]{fuchs}
     Fuchs, Y., Rodriguez, J., Mirabel, I.F., et al. 2003, A\&A, 409, L35

\bibitem{g99}
Gierli\'nski, M., Zdziarski, A.A., Poutanen, J., et al.\ 1999, \mnras, 309, 496


   \bibitem[Greiner et al. 2001]{greiner}
        Greiner, J., Cuby, J.G., McCaughrean, M.J., Castro-Tirado, A.,
        \& Mennickent, R.E. 2001, A\&A, 373, L37


\bibitem[Hannikainen et al. (2003)]{hannikainen}
      Hannikainen, D.C., Vilhu, O., Rodriguez, J., et al. 2003, A\&A,
      411, L415 

\bibitem[Harlaftis \& Greiner (2004)]{harlaftis}
        Harlaftis, E., \& Greiner, J. 2004, A\&A, 414, L13

\bibitem[Knodlseder et al. 2004]{knodelseder}
       Kn\"odlseder, J., et al. 2004, 
     Proceedings of the 5th INTEGRAL Workshop, Munich, Feb 16--20, 
     2004, ESA-SP 552.

\bibitem[Lebrun et al. 2003]{lebrun}
         Lebrun, F., Leroy, J.P., Lavocat, P., et al. 2003, A\&A, 411, L141

\bibitem[Lund et al. 2003]{lund}
        Lund, N., Budtz-J{\o}rgensen, C., Westergaard, N.J., et al. 2003,
        A\&A, 411, L231

\bibitem{mz95}
Magdziarz, P., \& Zdziarski, A.A. 1995, MNRAS, 273, 837

\bibitem[Mas-Hesse et al. 2003]{mh03}
   Mas-Hesse, J.M., A. Gim\'enez, J.L. Culhane, et al. 2003, A\&A,
   411, L261

\bibitem[Merloni et al. 2000]{merloni}
       Merloni, A., Fabian, A.C., \& Ross, R.R. 2000, MNRAS, 313, 193
 
\bibitem[Mitsuda et al. 1984]{mitsuda}
       Mitsuda, K., Inoue, H., Koyama, K., et al. 1984, PASJ, 36, 741 

\bibitem[Miyamoto et al. 2003]{miyamoto}
        Miyamoto, S., Kitamoto, S., Hayashida, K., \& Egoshi, W.,
        1995, ApJ, 442, L13

\bibitem[Morgan et al. 1997]{morgan}
          Morgan, E.H., Remillard, R.A., \& Greiner, J. 1997, ApJ, 482, 993

\bibitem[Poutanen \& Coppi 1998]{pc98}
     Poutanen, J., \& Coppi, P. 1998, Physica Scripta, T77, 57

\bibitem[Poutanen \& Svensson 1996]{ps}
    Poutanen, J., \& Svensson, R. 1996, ApJ, 470, 249

\bibitem[Poutanen et al. 1997]{poutanen}
    Poutanen, J., Krolik, J.H., \& Ryde, F. 1997, MNRAS, 292, L21


\bibitem[Rodriguez et al. 2002]{rodriguez02}
          Rodriguez, J., Varni\`ere, P., Tagger, M., \& Durouchoux, P. 
          2002, A\&A, 387, 487

\bibitem[Rodriguez et al. 2003]{rodriguez03}
    Rodriguez, J., Corbel, S., \& Tomsick, J.A. 2003, ApJ, 595, 1032

\bibitem[Rodriguez et al. (2004a)]{rodriguez2004b}
     Rodriguez, J., Corbel, S., Hannikainen, D.C., et al.
     2004a, ApJ, 615, 416

\bibitem[Rodriguez et al. (2004b)]{rodriguez2004b}
     Rodriguez, J., Cabanac, C., Hannikainen, et al., 2004b, astro-ph/0412555

\bibitem[Rodriguez \& Mirabel]{rm}
  Rodr\'\i guez, L.F., \& Mirabel, I.F. 1999, ApJ, 511, 398

\bibitem[Shimura \& Takahara 1995]{shimura}
    Shimura, T., \& Takahara, F. 1995, ApJ,445, 780

\bibitem[Skinner \& McConnell 2003]{sc}
  Skinner, G., \& McConnell, P. 2003, A\&A, 411, L123

\bibitem{sz03}
Sobolewska, M.A., \& \.Zycki, P. T. 2003, A\&A, 400, 553
 
\bibitem[Tagger et al. 2004]{tagger}
     Tagger, M., Varni\`ere, P., Rodriguez, J., \& Pellat, R. 2004,
     ApJ, 607, 410

\bibitem[Ubertini et al. 2003]{ubertini}
    Ubertini, P., Lebrun, F., Di Cocco, G., et al. 2003, A\&A, 411, L131

\bibitem[Vedrenne et al. 2003]{vedrenne}
    Vedrenne, G., Roques, J.P., Schoenfelder, V., et al. 2003, A\&A,
    411, L63

\bibitem[Vilhu \& Nevalainen 1998]{vilhu}
    Vilhu, O., \& Nevalainen, J. 1998, ApJ, 508, L85 

\bibitem[Vilhu]{vilhu01}
    Vilhu, O., Poutanen, J., Nikula, P., \& Nevalainen, J. 2001, ApJ,
    553, L51

\bibitem[vilhu03]{vilhu03}
    Vilhu, O., Hjalmarsdotter, L., Zdziarski, A.A., et al. 2003, A\&A,
    411, L405

\bibitem[Winkler et al. (2003)]{winkler}
      Winkler, C., Courvoisier, T.J.-L., DiCocco, G., et al. 2003, 
      A\&A, 411, L1

   \bibitem[Zdziarski et al. (2001)]{zdziarski}
       Zdziarski, A.A., Grove, J.E., Poutanen, J., Rao, A.R., \&
       Vadawale, S.V. 2001, ApJ, 554, L45

\bibitem{z04}
Zdziarski, A.A., Gierli\'nski, M., Miko{\l}ajewska, J., et al. 2004,
MNRAS, 351, 791



\end{thebibliography}
\end{document}